# Harnessing Modal Fields Retrieved from Speckle for Multi-Dimensional Metrology

*Qingbo Liu, Zhongyang Xu,\* Guangkui Tao, Xiuyuan Sun, Min Xue, Weihao Yuan, and Shilong Pan\**

Qingbo Liu. Zhongyang Xu. Author 1
Guangkui Tao. Author 2
Xiuyuan Sun. Min Xue. Weihao Yuan. Shilong Pan. Author 3
Key Laboratory of Microwave Photonics, College of Electronic and Information Engineering, Nanjing University of Aeronautics and Astronautics, Nanjing 211106, China.
E-mail: xzy@nuaa.edu.cn; pans@nuaa.edu.cn

*Abstract*

Although speckle is a powerful tool for high-precision metrology, large datasets and cumbersome training are always required to learn from the encoded speckle patterns, which is unfavorable for rapid deployment and multi-dimensional metrology. To enable high accuracy and fast training, physics-informed machine learning enforces physical laws to address high-dimensional problems. Here, we harness the modal fields in a few-mode fiber, which follow the law of beam propagation, to enable high-accuracy and fast-training parameter estimation. Anti-noise fast mode decomposition is implemented to retrieve the modal fields from the speckles. The accuracy is enhanced since the modal fields enable parameter estimation at random points in the continuous space-time domain. Artificial tactile perception and multi-dimensional metrology are achieved with high accuracy because the modal fields respond diversely to different parameters. Meanwhile, the number of specklegrams for training is reduced by around 5 times. The training time of machine learning is significantly reduced by 800 times, from 9 hours and 45 minutes to 40 seconds. Therefore, harnessing the modal fields paves a new way for the speckle-based metrology to develop efficient, low-cost, multi-dimensional sensors, making it suitable for intelligent wearable devices, industrial robots and healthcare applications.

## 1. Introduction

Metrology and sensing serve as the bridges connecting the physical and digital worlds. Multi-dimensional metrology and sensing, which estimate multiple parameters at the same time, significantly enhance the perception capabilities of intelligent devices,[1–3] demonstrating immense applications in smart wearable devices,[4–7] intelligent robots[8–12] and healthcare.[13–16] Speckle is a powerful tool for high-precision metrology and sensing,[17–19] since it enables remarkable performance in diverse measurement scenarios with simple setups. By incorporating machine learning, speckle-based metrology and sensing can achieve measurement ranges that exceed traditional linear response intervals,[20–23] without the need for human prior knowledge. However, existing approaches are typically enabled by data-driven machine learning. The performance is limited by the amount of training data and sophisticated models.[24–31] In classification tasks, only specific states and values can be estimated from the specklegrams, which perform poorly when handling continuously varying parameters.[24,27–31] In regression tasks, estimation accuracy relies heavily on deep models, which entails great computational cost.[25,26] Consequently, to achieve multi-dimensional metrology and sensing with high accuracy, cumbersome training with a large amount of specklegram data and sophisticated models is required to estimate the parameters from the speckle patterns. Typically, training a deep learning network demands dozens of hours,[32] which remains more than one hour even after optimizing the algorithm.[33]

Instead, physics-informed machine learning trains models using additional information obtained by enforcing the physical laws.[34–36] It integrates noisy data and physical models, enabling parameter estimation at arbitrary points in the continuous space-time domain. Hence, higher accuracy, faster training and improved generalization are expected.[34] As for the speckle in a few-mode fiber (FMF), it arises from the random interference of distinct modal fields. Leveraging the modal fields contained in speckle holds promise for addressing high-dimensional problems. Different from the data-driven approaches, modal fields follow the physical laws of beam propagation in fiber.[37] The significance of harnessing the modal fields lies in that it can achieve higher accuracy without relying on deeper models. Consequently, the training time and data volume are largely reduced. Moreover, multiple modes respond diversely to different parameters, which provides additional dimensions for multi-dimensional metrology and sensing.[38,39]

The primary premise of harnessing the modal fields is to obtain them in the fiber. Mode decomposition is required to accurately retrieve the amplitude and phase of each mode from speckles. Existing approaches can be classified into two categories. The first one based on experimental metrology[40–44] provides accurate and intuitive results. However, complex operational procedures and stringent experimental conditions restrict its applications. Another involves numerical analysis. It includes various iterative optimization,[45–48] such as stochastic parallel gradient descent (SPGD)[47] and genetic algorithm,[48] as well as non-iterative algorithms[49–54] like fractional Fourier transform system,[49] inverse matrix solution (IMS)[51] and 2D least-squares method.[52] The SPGD algorithm requires only a single intensity distribution of speckles to achieve mode decomposition. But the drawback is that it is susceptible to unexpected initial values and takes a long time. Although the time cost of the non-iterative IMS algorithm can be reduced to tens of microseconds, it is sensitive to noise, making it suitable for optical communications that are robust to noise. But for metrology and sensing, mode decomposition with short time and high accuracy in the presence of noise is demanded.

In this work, we harness the modal fields retrieved from speckle to enable high-accuracy and fast-training parameter estimation. The modal fields of an FMF are obtained from specklegrams using an anti-noise fast mode decomposition based on the law of beam propagation. Subsequently, mode-informed machine learning is employed for parameter estimation, including fiber curvature, bending position, bending angle and torsion angle. Proof-of-concept experiments confirm the enhanced accuracy. 2D artificial tactile perception of arbitrary patterns is also achieved. The number of specklegrams for training is reduced by 5 times. The training time of machine learning is significantly reduced by 800 times, from 9 hours and 45 minutes to 40 seconds. Moreover, modal fields allow multi-dimensional parameter estimation with low crosstalk, which enables high-accuracy and multi-dimensional metrology. Therefore, harnessing the modal fields paves a new way for the speckle-based metrology to develop efficient, low-cost, multi-dimensional sensors, making it suitable for intelligent wearable devices, industrial robots and healthcare applications.

## 2. Results and discussion

### 2.1. Concept of the metrology informed by modal fields

In an FMF, the optical field is a linear superposition of fiber eigenmodes. The optical intensity can be expressed as

$$I(x,y) = \left| \sum_{j=1}^{N} \rho_j e^{i\theta_j} \psi_j(x,y) \right|^2 \tag{1}$$

where $N$ is the number of eigenmodes, $\rho_j$ and $\theta_j$ are the normalized amplitudes and relative phases of eigenmodes, satisfying $\rho_j \in [0, 1]$ and $\theta_j \in [-\pi, \pi]$, respectively. $\Psi_j(x, y)$ can be described as the normalized field distribution of the $j^{th}$ order linear polarization (LP) mode.

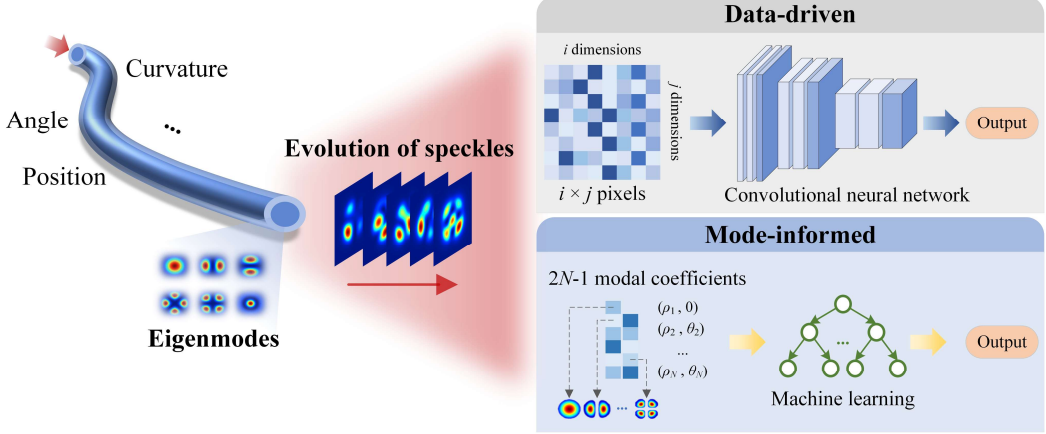

**Figure 1.** Schematic illustration of the speckle-based metrology and sensing using an FMF. The conventional approach is data-driven, while the proposed approach is informed by modal fields.

The concept of speckle-based metrology and sensing using an FMF is illustrated in **Figure 1**. Existing approaches are primarily enabled by data-driven machine learning, which demands a large amount of labeled specklegrams and time for training. Nevertheless, our approach utilizes the amplitudes and relative phases of $N$ eigenmodes, which is called mode-informed. It is the $2N-1$ modal coefficients rather than $i \times j$ pixel speckles that are harnessed for training the machine learning model. In the FMF, $N$ is always less than 10, whereas the number of pixels is generally more than $256 \times 256$. Hence, less time and computational cost are anticipated for the training informed by modal fields. Furthermore, the speckle patterns always change randomly while the modal fields are usually continuous and predictable. It may promise a higher-accuracy parameter estimation with a smaller amount of training data.

**2.2. Mode decomposition from speckle**

To harness the modal fields for speckle-based metrology and sensing, mode decomposition needs to be implemented. It involves extracting the $2N-1$ modal coefficients from the specklegrams, including the amplitudes and relative phases (with $\theta_1=0$). The intensity of optical field in the FMF can also be expressed as the product of the eigenmode matrix $H$ and the amplitude-phase vector $X$.

$$HX = I \qquad (2)$$

where

$$X = \begin{pmatrix} \rho_1^2 & \cdots & \rho_N^2 & \rho_2\rho_1\cos(\theta_2) & \cdots & \rho_N\rho_1\cos(\theta_N) & \rho_3\rho_2\cos(\theta_3-\theta_2) & \cdots & \rho_N\rho_{N-1}\cos(\theta_N-\theta_{N-1}) \end{pmatrix}^T \qquad (3)$$

According to Equation (3), the modal coefficients including $N$ modal amplitudes and $N-1$ phases are calculated.

$$\begin{cases} \rho_i = \sqrt{X_i}, \quad i = 1\cdots N \\ \theta_j = \arccos(\dfrac{X_{j+N-1}}{\rho_j\rho_1}), \quad j = 2\cdots N \\ \theta_j = \theta_j \cdot \mathbf{sgn}(|\rho_j\rho_2\cos(\theta_j+\theta_2) - X_{j+2N-3}| - |\rho_j\rho_2\cos(\theta_j-\theta_2) - X_{j+2N-3}|), \quad j = 3\cdots N \end{cases} \qquad (4)$$

where **sgn** represents the sign function and $\theta_2 \in [0, \pi]$.

Fast mode decomposition has been achieved using the IMS method.[51] Although its time consumption is quite low, the accuracy of decomposition is insufficient for high-accuracy metrology and sensing due to the system noise. Here, an improved mode decomposition with anti-noise capability is proposed, which includes preparatory stage and mode decomposition stage. In the preparatory stage, multiple specklegrams are collected and decomposed using the classic SPGD algorithm. The calculated modal coefficients are used to obtain the eigenmode matrix ($H'$) in the noisy environment. Then, the mapping between the specklegrams and the modal coefficients can be established. It should be clarified that the preparatory stage can be completed in one go without the need for repetition. In the mode decomposition stage, the modal coefficients are initially computed using the Equation $X=(H^2)^{-1}I$ and (4), which are further optimized to obtain precise values (see Section 1 in Supporting Information). Its noise robustness can be confirmed by simulations (see Section 2 in Supporting Information). Our method achieves a higher decomposition accuracy than the IMS and SPGD methods under the same signal-to-noise ratio (SNR) and number of modes. This advantage becomes more pronounced as the SNR decreases or the number of modes increases.

**Figure 2** illustrates the experimental evaluation of IMS, SPGD and our method. For 3,500 samples, the average correlations between the measured and the reconstructed specklegrams are 0.801, 0.979 and 0.985, respectively, as shown in Figure 2a. The standard deviations are 0.029, 0.006 and 0.004, respectively. Moreover, Figure 2b illustrates the performance at different SNRs. The increase in noise leads to large errors and a decrease of 0.07 in correlation for the IMS method. In contrast, the correlation for our method and SPGD only decreases by 0.001, indicating the excellent noise robustness. Consequently, the proposed method exhibits higher decomposition accuracy and stability in a noisy environment. Furthermore, it requires only 0.42 s to achieve single mode decomposition in a six-mode fiber, which is more than 10 times faster than the classic SPGD method. The time cost can be further reduced to 0.26 s if the sensing accuracy is lowered to the level of existing methods. Although our method is still slower than IMS, it is sufficient for fiber sensing and represents the optimal choice that balances decomposition accuracy and speed (see Section 5 in Supporting Information).

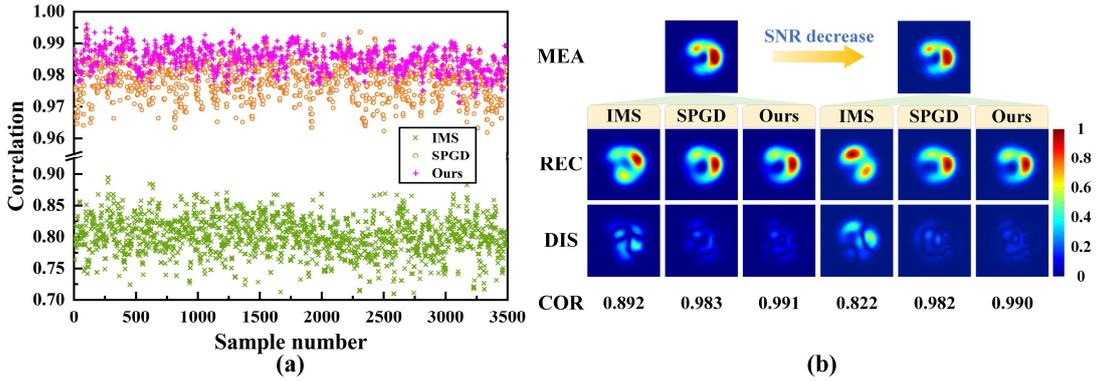

**Figure 2.** Experimental results of mode decomposition using three different algorithms (IMS, SPGD and our method). (a) Reconstructed correlations on 3,500 specklegrams. (b) Impact of noise on the mode decomposition. MEA: measured profile; REC: reconstructed profile; DIS: profile difference; COR: correlation.

**2.3. Metrology and sensing informed by modal fields**

The experimental setup for fiber sensing informed by modal fields is shown in **Figure 3**. To reduce the processing time and memory consumption without degrading estimation accuracy (see Section 3 in Supporting Information), we utilize only the amplitude information, which is the modal weight $\rho^2$, for sensing. The FMF is a standard six-mode commercial step-index fiber. A machine learning model based on the LightGBM algorithm[55] is trained to analyze the complex relationships between the modal weights and external sensing parameters. **Figure 4** illustrates the results of curvature sensing. It is significant that the modal weights vary continuously with curvature in Figure 4a, which is preferable for machine learning to achieve higher regression accuracy. Moreover, different modes exhibit different responses to the changes in curvature. Modes $LP_{01}$, $LP_{02}$ and $LP_{21o}$ show a strong correlation with curvature changes, while other modes exhibit weaker correlations.

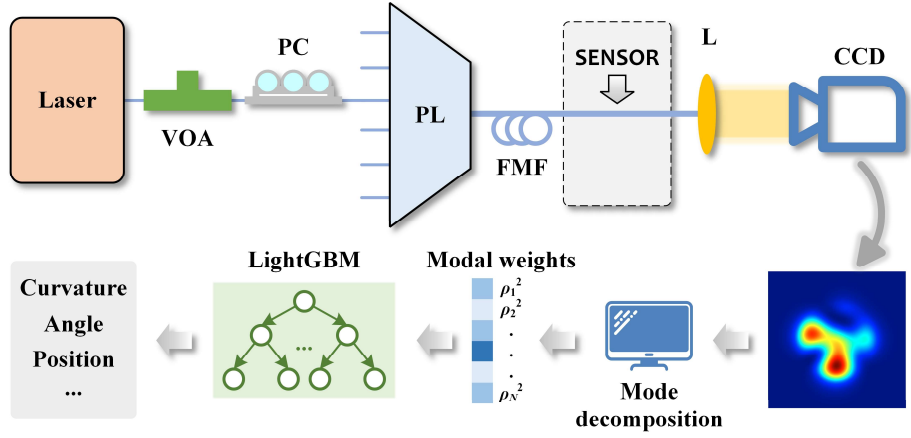

**Figure 3.** The experimental system and workflow for mode-informed metrology and sensing. VOA: variable optical attenuator; PC: polarization controller; PL: photonic lantern; L: lens.

In Figure 4c, the root mean square error (RMSE) on the validation set decreases to 0.032 m$^{-1}$ after 1,000 training iterations, close to that of the training set. It indicates that the model is effectively trained and possesses strong generalization capability. The performance of curvature sensing in the range from 0 to 3.83 m$^{-1}$ is shown in Figure 4d. The R² value for the estimated results is 0.9987. Figure 4e illustrates the estimation of step-change curvature, in which the standard deviation of the estimation errors is 0.033 m$^{-1}$. Moreover, multiple parameters, including bending position, bending angle and torsion angle, are also estimated and presented in **Figure 5a–c**. The R² values of the estimated results for bending position ($D$), bending angle ($\theta_1$) and torsion angle ($\theta_2$) are 0.9999, 0.9963 and 0.9933, respectively. The sensing ranges for them are 17 mm, 25° and 60°, respectively. The standard deviations of estimation errors for these three parameters are 0.041 mm, 0.502° and 1.402°, respectively (Figure 5d–f).

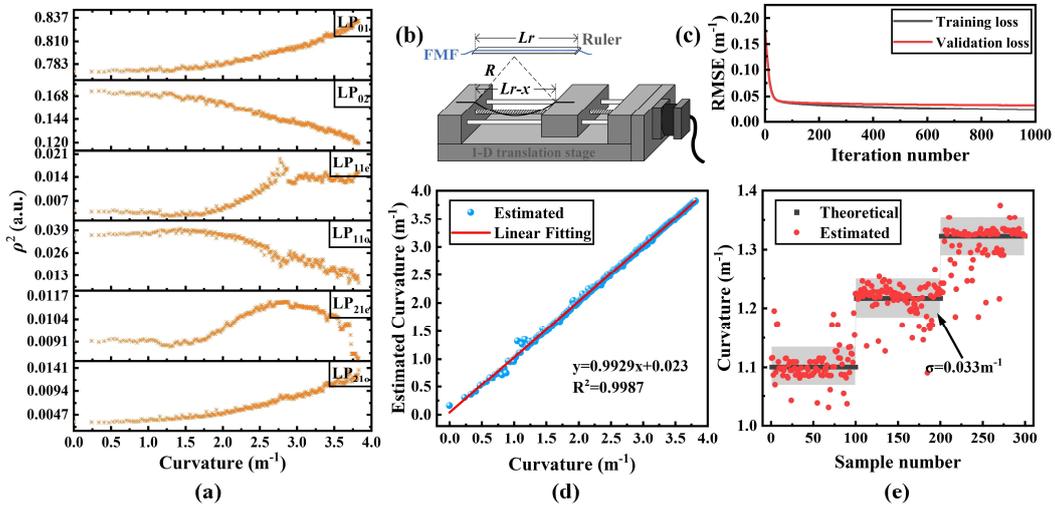

**Figure 4.** Curvature sensing informed by modal fields. (a) Evolution of modal weights with fiber curvature. (b) Experimental setup for changing fiber curvature. (c) Training and validation losses of machine learning. (d) Relationship between the estimated and actual curvature. (e) Distribution of estimated curvature for step changes.

**Table 1** summarizes the performance of existing speckle-based fiber sensors, including FMF, multimode fiber (MMF) and ring-core fiber (RCF) sensors. The resolution of conventional speckle sensors is limited by the number of discrete label classifications and model complexity since the digital specklegrams are sparse and vary randomly. Therefore, complex deep models are essential to ensure the accuracy. In our method, harnessing modal fields enables prediction of parameters over a large range with higher regression accuracy, since the parameters at any state can be estimated based on the continuous modal coefficient vectors. Compared to the current state-of-the-art speckle sensors, the measurement resolution in fiber curvature and bending angle sensing is enhanced by more than 3 times. The position resolution is improved by 7 times to 41 micrometers. Additionally, we also demonstrate axial rotation angle sensing on an FMF.

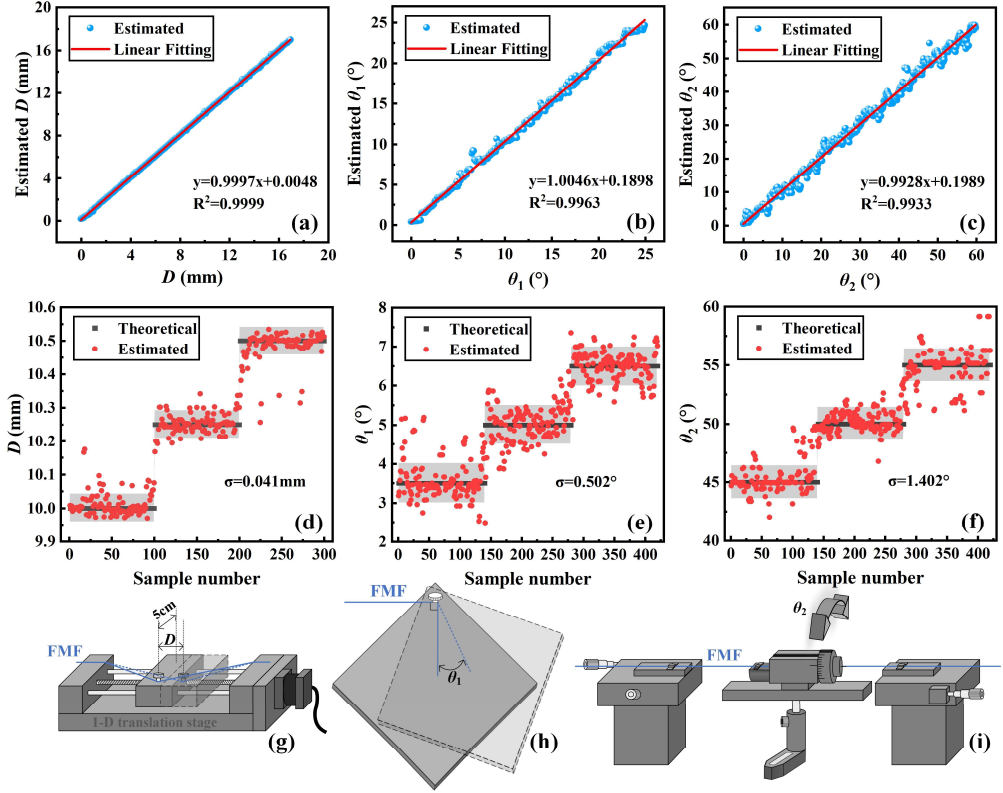

**Figure 5.** Bending position, bending angle and torsion angle sensing informed by modal fields. (a)–(c) Relationship between the estimated values and the actual values. (d)–(f) Distribution of estimated parameters for step changes. (g)–(i) Experimental setups for changing the bending position, bending angle and torsion angle, respectively.

Table 1. Various types of speckle-based fiber sensors.

| Ref | Structure | Method | Task | Type | Measurement Range | Resolution |
|---|---|---|---|---|---|---|
| Ref. [24] | MMF | CNN | Curvature | Classification | 1.23~5.46 m$^{-1}$ | 0.202 m$^{-1}$ |
| Ref. [25] | MMF | CNN | Curvature | Regression | 1.55~6.93 m$^{-1}$ | 0.1 m$^{-1}$ |
| Ref. [26] | MMF | CNN | Bending | Regression | 15°~345° | 30° |
| Ref. [27] | MMF | KNN | Bending | Classification | 88.2° | 1.8° |
| | | | Position | | 6 mm | 0.3 mm |
| Ref. [28] | RCF | CNN | Position | Classification | 50 cm | 5 cm |
| Ref. [29] | MMF | CNN | Position | Classification | 8 mm (4D) | 0.8 mm |
| Our work | FMF | LightGBM | Curvature | Regression | 0~3.83 m$^{-1}$ | 0.033 m$^{-1}$ |
| | | | Bending | | 25° | 0.502° |
| | | | Torsion | | 60° | 1.402° |
| | | | Position | | 17 mm | 0.041 mm |

The primary novelty of the method lies in leveraging modal fields instead of directly employing random specklegrams. Since the modal fields are continuous and predictable, high-accuracy estimation can be expected. Furthermore, mode decomposition refines the encoded speckle containing thousands of pixels into several modal coefficients, avoiding the direct handling of cumbersome specklegrams. Hence, training model requires less time and memory resources. We additionally construct a convolutional neural network (CNN) that directly processes specklegrams on the same computing platform. Both models are trained to perform regression tasks on a six-mode fiber speckle dataset, where the specklegram size is 256 × 256 pixels. The CNN model is trained on 20,000 speckle pattern samples, while our approach based on modal fields achieves comparable estimation accuracy with only 4,200 samples, saving nearly 5 times. These operations are conducted on an Intel i7-11800H CPU and an NVIDIA RTX 3060 GPU. In **Table 2**, the training time for the model is reduced by over 800 times, to 40 s, while it takes 9 hours and 45 minutes directly using specklegrams. For inference time, our scheme requires only 0.015 ms, reduced by around 100 times. It should be clarified that the total time cost should include the time for mode decomposition and model training, which is around 20 minutes, reduced by 30 times. Consequently, the reduction in training time and memory consumption renders our method suitable for rapid deployment in varied surroundings.

**Table 2.** Time consumption of speckle sensors.

| Sensing schemes | Training time | Inference time |
| --- | --- | --- |
| Specklegrams | ~9 h 45 min | ~1.4 ms |
| Modal weights (Ours) | ~40 s | ~0.015 ms |

**2.4. Tactile sensation and multi-dimensional metrology**

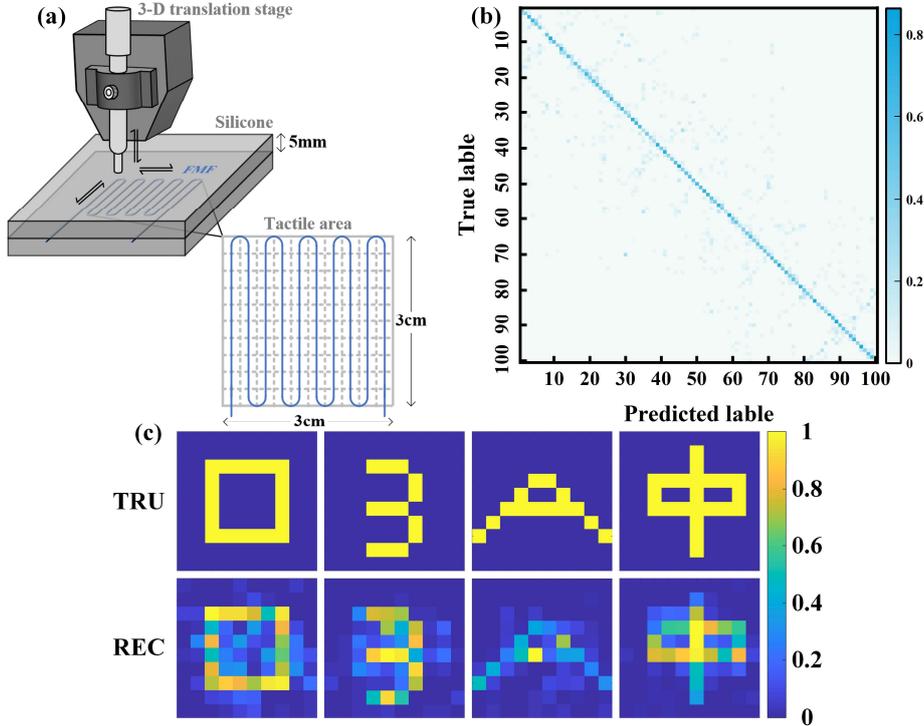

**Figure 6.** Tactile sensing informed by modal fields. (a) Experimental setup for tactile sensation. (b) Confusion matrix for the predictions of the 10 × 10 pixel positions. (c) Actual images (TRU) of four different tactile patterns and corresponding reconstructed results (REC).

Since the sensor enables stable and high-accuracy metrology, an intelligent tactile sensor can be achieved. **Figure 6a** shows the experimental setup for tactile sensation, in which a six-mode fiber is clamped between two flexible silicone plates and arranged in a uniform S-shape. A metal pen with a 2-mm contact point draws arbitrary patterns on a 3 cm × 3 cm silicone plate. The plate is divided into 10 × 10 pixels, yielding a position resolution of 3 mm × 3 mm. Modal weights are utilized to estimate the bending positions, thereby reconstructing arbitrary patterns and enabling tactile sensation. For the validation set, the confusion matrix of predicting the position of all pixels is shown in Figure 6b. Then, four patterns are drawn, including a square, number, letter and Chinese character. Figure 6c displays the reconstructed results, where all four patterns are effectively reconstructed. The proposed 2D tactile sensor demonstrates superior performance in sensing range and stability compared to existing optical tactile sensing schemes (see Section 4 in Supporting Information). This is primarily attributed to the utilization of modal information that reduces the impact of speckle noise. It is crucial for achieving high-resolution tactile sensation while maintaining a large measurement range.

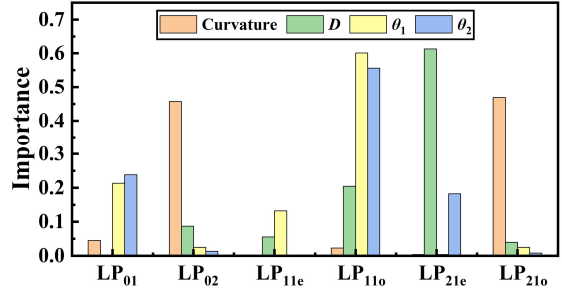

**Figure 7.** Importance of multiple modes to fiber curvature, bending position, bending angle and torsion angle sensing in a six-mode optical fiber.

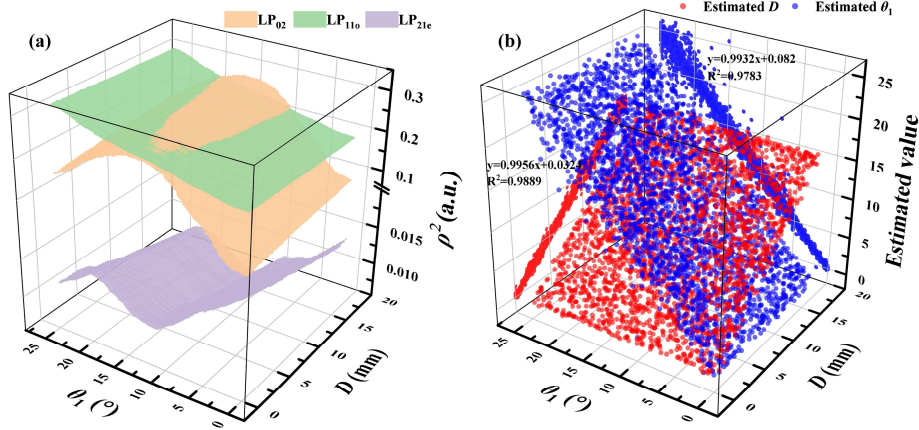

**Figure 8.** Dual-parameter estimation with an FMF. (a) Evolution of modal weights ($LP_{02}$, $LP_{11o}$ and $LP_{21e}$) with the bending position $D$ and bending angle $\theta_1$. (b) Simultaneous dual-parameter estimation.

Another attractive feature of the mode-informed speckle sensor is that different modes can be considered as different eigenspaces, which exhibit diverse responses to external parameters. This forms the foundation for multi-dimensional parameter estimation. **Figure 7** illustrates the importance of modal weights in estimating fiber curvature, bending position, bending angle and torsion angle. A higher importance indicates a greater information gain for the estimated results and a stronger correlation between the modal weights and external parameters. For curvature, the importances of $LP_{01}$, $LP_{02}$ and $LP_{21o}$ are relatively high. This is because fiber bending leads to uneven changes in radial refractive index. For the torsion angle, the changes in refractive index occur around the fiber axis, which may explain the greater importances of the $LP_{01}$, $LP_{11o}$ and $LP_{21e}$ modes. **Figure 8a** shows how modal weights vary with changes in bending position and bending angle. The modal weights of $LP_{02}$ and $LP_{21e}$ exhibit a stronger linear correlation with the bending position, while $LP_{11o}$ mode shows greater variation across different bending angles.

These features promise multi-dimensional sensing of various parameters. The estimated results for bending position and angle are shown in Figure 8b. The bending position and angle at any state can be simultaneously estimated. The $R^2$ values for the projection data are 0.9889 and 0.9783, respectively. Compared to single-parameter sensing, the $R^2$ values in dual-parameter sensing are slightly lower, reflecting the increased estimation error that may arise from the cross-talk of coupled data, highlighting the trade-off between sensing accuracy and the number of parameters.

## 3. Conclusion

Although renowned speckle-based metrology and sensing detect minute variations through simple device deployment, the actual performance heavily depends on large datasets and deep models. In this study, we implement a concept of harnessing modal fields retrieved from speckle for multi-dimensional metrology using an FMF. It is the physical information of optical fields, rather than the digital specklegrams containing thousands of pixels, that is used for parameter estimation. An anti-noise mode decomposition is proposed to enable fast and accurate extraction of each modal coefficient. It shows significant advantages in accuracy and stability, with a consumption time over an order of magnitude lower than that of SPGD. The excellent mode decomposition can also capture nonlinear effects in transient intermodal interactions, facilitating the exploration of potential excitation mechanisms, dispersion control,[56] and mode manipulation dynamics.

Metrology and sensing informed by modal fields have achieved better performance. The accuracy of curvature and bending angle is enhanced by 3 times, and 7 times for bending position. Meanwhile, harnessing modal fields in the speckle sensor reduces training dataset by 5 times. It arises from the effective training of modes that follow the beam propagation model, instead of being wasted on learning speckle noise. The inference is accelerated by 100 times and the training time is significantly reduced by over 800 times, requiring only 40 s. All these avoid the excessive dependence on complex networks and large datasets, reducing computational consumption and calibration time. We also demonstrate 2D tactile perception, where arbitrary patterns can be well reconstructed without prior knowledge. Additionally, various parameters impact the optical modes in different ways by altering the refractive index distribution. Modal domain allows for the simultaneous estimation of bending position and angle with low crosstalk. With an elaborate design, the combination of several modes with different weights can further enhance the estimation accuracy. This study explores the potential of measurement in modal domain. We anticipate that it will pave the way for speckle-based metrology to develop efficient, low-cost, multi-dimensional sensors that are suitable for intelligent wearable devices, as well as human-robot and robot-environmental interactions.[57]

## 4. Experimental Section

**Mode decomposition**

A laser emits light at 1550 nm, which enters a variable optical attenuator (VOA) and a polarization controller (PC). The VOA adjusts the optical power, while the PC selects the polarization of the beam. Then, the light is coupled into FMF using a six-mode photonic lantern (PL). The FMF is a standard six-mode commercial step-index fiber (FM2011-B, YOFC) with a length of 10 m. It has a core diameter of 18.5 μm and a cladding diameter of 125 μm, supporting six LP modes: $LP_{01}$, $LP_{02}$, $LP_{11e}$, $LP_{11o}$, $LP_{21e}$ and $LP_{21o}$. The beam profile from FMF is captured by a CCD camera (Bobcat-320-GigE, Xenics) with a pixel spacing of 20 μm. All processed images have a pixel size of 256 × 256. The initial modal coefficients calculated from Equation (4) are further iterated rapidly to obtain accurate decomposition results using the Broyden-Fletcher-Goldfarb-Shanno (BFGS) optimization algorithm.[58]

**Multi-parameter metrology**

To change the curvature, a fiber is attached along a ruler (Figure 4b). The translation stage bends the ruler by squeezing it, which alters the curvature[59] within the range of 0 to 3.83 $m^{-1}$. To change the bending position, a 40 cm-long FMF is fixed at both ends to the platform (Figure 5g). A steel rod is used to pull the fiber in the perpendicular direction. The rod moves with translation stage, causing the fiber to bend at different positions.[27] To change the bending angle, one end of the fiber is fixed while the other end bends in a plane at different angles (Figure 5h). To change the torsion angle, a 28 cm-long FMF, secured horizontally by a fiber clamp, is twisted around its axial direction using a fiber rotator (Figure 5i). Tactile actions are implemented through a metal pen (Figure 6a), which is controlled by a 3-D translation stage to sequentially press each position. The pressing depth is maintained at 4 mm to avoid estimation errors caused by pressure variations before and after model training. Machine learning based on the LightGBM algorithm is used to analyze the complex relationship between the modal fields and external parameters.

**Supporting Information**

Supporting Information is available from the Wiley Online Library or from the author.

**Acknowledgements**

This project was supported by the National Natural Science Foundation of China (62171219, 62271249, 62071226, 62401257).

**Conflict of Interest**

The authors declare no conflict of interest.

**Author Contributions**

Q.L. and Z.X. contributed equally to this work. Z.X. initiated the study and designed the experiments. Q.L. conducted the machine learning and carried out the experiments. G.T. carried out the experiments and analyzed the data. M.X. and W.Y. guided the experiments. Z.X., Q.L. and S.P. wrote the paper. Z.X. and S.P. supervised the project.

**Data Availability Statement**

The data that support the findings of this study are available from the corresponding author upon reasonable request.

**References**

[1] N. Lalam, S. Bukka, H. Bhatta, M. Buric, P. Ohodnicki, R. Wright, *Commun. Eng.* **2024**, *3*, 121.

[2] M. Chen, Z. Wang, Q. Zhang, Z. Wang, W. Liu, M. Chen, L. Wei, *Nat. Commun.* **2021**, *12*, 1416.

[3] R. Wang, R. Wang, C. Dou, S. Yang, R. Gnanasambandam, A. Wang, Z. Kong, *Nat. Commun.* **2024**, *15*, 7568.

[4] Y. Li, Y. Su, *Light Sci. Appl.* **2024**, *13*, 79.

[5] P. Li, Y. Wang, X. He, Y. Cui, J. Ouyang, J. Ouyang, Z. He, J. Hu, X. Liu, H. Wei, Y. Wang, X. Lu, Q. Ji, X. Cai, L. Liu, C. Hou, N. Zhou, S. Pan, X. Wang, H. Zhou, C.-W. Qiu, Y.-Q. Lu, G. Tao, *Light Sci. Appl.* **2024**, *13*, 48.

[6] J. Guo, M. Niu, C. Yang, *Optica* **2017**, *4*, 1285.

[7] A. Leal-Junior, L. Avellar, A. Frizera, C. Marques, *Sci. Rep.* **2020**, *10*, 13867.

[8] Y. Li, Z. Guo, X. Zhao, S. Liu, Z. Chen, W.-F. Dong, S. Wang, Y.-L. Sun, X. Wu, *Nat. Commun.* **2024**, *15*, 2906.

[9] Z. Peng, H. Wen, J. Jian, A. Gribok, M. Wang, S. Huang, H. Liu, Z.-H. Mao, K. P. Chen, *Sci. Rep.* **2020**, *10*, 21014.

[10] L. Li, F. Yang, Q. Ma, T. Xu, M. Su, Y. Zhao, X. Liu, *Optica* **2025**, *12*, 263.

[11] H. C. Lee, N. Elder, M. Leal, S. Stantial, E. Vergara Martinez, S. Jos, H. Cho, S. Russo, *Nat. Commun.* **2024**, *15*, 8456.

[12] D. Feng, W. Zhou, X. Qiao, J. Albert, *Sci. Rep.* **2015**, *5*, 17415.

[13] S.-M. Jeong, M. Son, Y. Kang, J. Yang, T. Lim, S. Ju, *Npj Flex. Electron.* **2021**, *5*, 35.

[14] Z. Che, X. Wan, J. Xu, C. Duan, T. Zheng, J. Chen, *Nat. Commun.* **2024**, *15*, 1873.

[15] D. Lo Presti, F. Santucci, C. Massaroni, D. Formica, R. Setola, E. Schena, *Sci. Rep.* **2021**, *11*, 21162.

[16] A. Issatayeva, A. Amantayeva, W. Blanc, D. Tosi, C. Molardi, *Sci. Rep.* **2021**, *11*, 8609.

[17] M. Facchin, S. N. Khan, K. Dholakia, G. D. Bruce, *Nat. Rev. Phys.* **2024**, *6*, 500.

[18] B. Lengenfelder, F. Mehari, M. Hohmann, M. Heinlein, E. Chelales, M. J. Waldner, F. Klämpfl, Z. Zalevsky, M. Schmidt, *Sci. Rep.* **2019**, *9*, 1057.

[19] S. Shimadera, K. Kitagawa, K. Sagehashi, Y. Miyajima, T. Niiyama, S. Sunada, *Sci. Rep.* **2022**, *12*, 13096.

[20] C. Lu, J. Su, X. Dong, T. Sun, K. T. V. Grattan, *J. Light. Technol.* **2018**, *36*, 2796.

[21] Y. Zhao, L. Cai, X.-G. Li, *IEEE Trans. Instrum. Meas.* **2017**, *66*, 141.

[22] X. Cai, S. Gao, M. Wu, S. Zheng, H. Fu, D. Chen, *Opt. Laser Technol.* **2024**, *168*, 109843.

[23] R. K. Gupta, G. D. Bruce, S. J. Powis, K. Dholakia, *Laser Photonics Rev.* **2020**, *14*, 2000120.

[24] Y. Liu, G. Li, Q. Qin, Z. Tan, M. Wang, F. Yan, *Opt. Laser Technol.* **2020**, *131*, 106424.

[25] G. Li, Y. Liu, Q. Qin, X. Zou, M. Wang, W. Ren, *IEEE Sens. J.* **2022**, *22*, 15974.

[26] S. Razmyar, M. T. Mostafavi, *J. Light. Technol.* **2021**, *39*, 1850.

[27] X. Wang, Y. Wang, K. Zhang, K. Althoefer, L. Su, *Sci. Rep.* **2022**, *12*, 12684.

[28] M. Wei, G. Tang, J. Liu, L. Zhu, J. Liu, C. Huang, J. Zhang, L. Shen, S. Yu, *J. Light. Technol.* **2021**, *39*, 6315.

[29] K. Sun, Z. Ding, Z. Zhang, *Appl. Opt.* **2020**, *59*, 5745.

[30] Z. Ding, Z. Zhang, *Opt. Laser Technol.* **2021**, *136*, 106760.

[31] X. Liu, S. He, J. Kang, B. Liu, C. Zhu, *Opt. Express* **2024**, *32*, 31783.

[32] Y. Li, Y. Xue, L. Tian, *Optica* **2018**, *5*, 1181.

[33] B. Rahmani, D. Loterie, G. Konstantinou, D. Psaltis, C. Moser, *Light Sci. Appl.* **2018**, *7*, 69.

[34] G. E. Karniadakis, I. G. Kevrekidis, L. Lu, P. Perdikaris, S. Wang, L. Yang, *Nat. Rev. Phys.* **2021**, *3*, 422.

[35] C. Zuo, J. Qian, S. Feng, W. Yin, Y. Li, P. Fan, J. Han, K. Qian, Q. Chen, *Light Sci. Appl.* **2022**, *11*, 39.

[36] X. Jiang, D. Wang, Q. Fan, M. Zhang, C. Lu, A. P. T. Lau, *Laser Photonics Rev.* **2022**, *16*, 2100483.

[37] Z. Tian, L. Pei, J. Wang, K. Hu, W. Xu, J. Zheng, J. Li, T. Ning, *Opt. Express* **2022**, *30*, 39932.

[38] X. Zhan, Y. Liu, M. Tang, L. Ma, R. Wang, L. Duan, L. Gan, C. Yang, W. Tong, S. Fu, D. Liu, Z. He, *Opt. Express* **2018**, *26*, 15332.

[39] I. Ashry, Y. Mao, A. Trichili, B. Wang, T. K. Ng, M.-S. Alouini, B. S. Ooi, *IEEE Access* **2020**, *8*, 179592.

[40] J. Demas, S. Ramachandran, *Opt. Express* **2014**, *22*, 23043.

[41] D. M. Nguyen, S. Blin, T. N. Nguyen, S. D. Le, L. Provino, M. Thual, T. Chartier, *Appl. Opt.* **2012**, *51*, 450.

[42] M. Paurisse, L. Lévèque, M. Hanna, F. Druon, P. Georges, *Opt. Express* **2012**, *20*, 4074.

[43] Y. Z. Ma, Y. Sych, G. Onishchukov, S. Ramachandran, U. Peschel, B. Schmauss, G. Leuchs, *Appl. Phys. B* **2009**, *96*, 345.

[44] A. Forbes, A. Dudley, M. McLaren, *Adv. Opt. Photonics* **2016**, *8*, 200.

[45] R. Brüning, P. Gelszinnis, C. Schulze, D. Flamm, M. Duparré, *Appl. Opt.* **2013**, *52*, 7769.

[46] O. Shapira, A. F. Abouraddy, J. D. Joannopoulos, Y. Fink, *Phys. Rev. Lett.* **2005**, *94*, 143902.

[47] H. Lü, P. Zhou, X. Wang, Z. Jiang, *Appl. Opt.* **2013**, *52*, 2905.

[48] L. Li, J. Leng, P. Zhou, J. Chen, *Opt. Express* **2017**, *25*, 19680.

[49] W. Yan, X. Xu, J. Wang, *Opt. Express* **2019**, *27*, 13871.

[50] E. Manuylovich, A. Donodin, S. Turitsyn, *Opt. Express* **2021**, *29*, 36769.

[51] E. S. Manuylovich, V. V. Dvoyrin, S. K. Turitsyn, *Nat. Commun.* **2020**, *11*, 5507.

[52] P. S. Anisimov, V. V. Zemlyakov, J. Gao, *Opt. Express* **2022**, *30*, 8804.

[53] Z. Tian, L. Pei, J. Wang, J. Zheng, K. Hu, Y. Chang, Z. Li, J. Li, T. Ning, *J. Light. Technol.* **2022**, *40*.

[54] Y. An, L. Huang, J. Li, J. Leng, L. Yang, P. Zhou, *Opt. Express* **2019**, *27*, 10127.

[55] K. Cheng, X. Yang, X. Wu, *IEEE Trans. Geosci. Remote Sens.* **2024**, *62*, 1.

[56] J. Ye, X. Ma, Y. Zhang, J. Xu, H. Zhang, T. Yao, J. Leng, P. Zhou, *PhotoniX* **2021**, *2*, 15.

[57] L. Wu, X. Yuan, Y. Tang, S. Wageh, O. A. Al-Hartomy, A. G. Al-Sehemi, J. Yang, Y. Xiang, H. Zhang, Y. Qin, *PhotoniX* **2023**, *4*, 15.

[58] J. Nocedal, S. J. Wright, *Numerical Optimization*, Springer, New York, **1999**.

[59] R. Falciai, C. Trono, *IEEE Sens. J.* **2005**, *5*, 1310.